\documentclass[prb,amsmath,amssymb,showpacs,showkeys,twocolumn]{revtex4}
\usepackage{graphicx}
\usepackage{dcolumn}
\usepackage{rotating}
\usepackage{amssymb}
\usepackage{mathptmx}

\usepackage{amsfonts}
\usepackage{amsmath}
\usepackage{bm}
\begin{document}

\title {Sliding of Electron Crystal of Finite Size on the Surface of Superfluid $^4$He Confined in a Microchannel}

\author{J.-Y. Lin}
\author{A.~V. Smorodin}
\altaffiliation[Current address: ]{Electronic and Quantum Magnetism Unit, Okinawa Institute of Science and Technology (OIST) Graduate University, Onna, 904-0495 Okinawa, Japan}
\author{A.~O. Badrutdinov}
\altaffiliation{Current address: Mechanical Engineering and Microfabrication Support Section, Okinawa Institute of Science and Technology (OIST) Graduate University, Onna, 904-0495 Okinawa, Japan}
\author{D. Konstantinov}
\email{denis@oist.jp}

\affiliation{Quantum Dynamics Unit, Okinawa Institute of Science and Technology (OIST) Graduate University, Onna, 904-0495 Okinawa, Japan}

\begin{abstract}

We present a new study of the nonlinear transport of a two-dimensional electron crystal on the surface of liquid helium confined in a $10$-$\mu$m-wide channel in which the effective length of the crystal can be varied from 10 to 215 $\mu$m. At low driving voltages, the moving electron crystal is strongly coupled to deformation of the liquid surface arising from resonant excitation of surface capillary waves, ripplons, while at higher driving voltages the crystal decouples from the deformation. We find strong dependence of the decoupling threshold of the driving electric field acting on the electrons, on the size of the crystal. In particular, the threshold electric field significantly decreases when the length of the crystal becomes shorter than 25~$\mu$m. We explain this effect as arising from weakening of surface deformations due to radiative loss of resonantly-excited ripplons from an electron crystal of finite size, and we account for the observed effect using an instructive analytical model.            

\end{abstract}

\date{\today}

c

\maketitle

\section{Introduction}

Electrons floating on the surface of liquid helium present a clean two-dimensional electron system with exquisitely well controlled parameters~\cite{Andrei}. For low electron densities of order $10^{13}$~m$^{-2}$ accessible on the surface of bulk liquid helium, the electrons form a classical non-degenerate system. On the other hand, the strong unscreened Coulomb interaction between electrons facilitates crystallization into a Wigner Solid (WS) phase, even at moderate cryogenic temperatures around 1~K. Study of the classical WS in this system, where it was experimentally realized for the first time,~\cite{Grim1979} complements studies of WS in a variety of other systems, including semiconductor heterostructures,~\cite{Andr1988,Zhu2010,Liu2014,Zhan2014} colloidal systems,~\cite{Murr1987,Murr1989,Murr1990,Marc1996} dusty plasma,~\cite{Chia1996} etc. In particular, interaction of WS with a soft substrate leads to peculiar nonlinear transport properties relevant, for example, to the general study of friction.~\cite{Vano2007,Wang2008,Vano2013,Byli2015} WS pressed against the liquid helium substrate causes a commensurate periodic deformation of the liquid surface called the dimple lattice. Coupling between WS and the dimple lattice leads to significant enhancement of the electron effective mass, thus altering the transport properties of the electron system as it is driven along the surface by an external electrical force. The most intriguing phenomenon in non-linear transport of WS is the saturation of electron velocity with the increasing driving force, which can be observed as a plateau in a measured $IV$-curve.~\cite{Kris1996,Ikeg2009}  Another striking phenomenon is the decoupling of WS from the dimples when the driving force on electrons exceeds some threshold value.~\cite{Shir1995,Rees2016} Such WS sliding from the dimples is accompanied by an abrupt increase in the measured current of electrons. 

An elegant explanation of the first phenomenon was given by Dykman and Rubo in terms of a coherent Bragg-Cherenkov (BC) emission of surface capillary waves, ripplons.~\cite{Dykm1997} As electron velocity $\textbf{v}$ approaches the phase velocity of ripplons, corresponding to their wave vector $\textbf{q}$, which is equal to the first reciprocal-lattice vector $\textbf{G}_1$ of WS, the emitted ripplons constructively interfere, increasing the dimple depth; therefore the effective electron mass. Theory predicts divergence of the frictional force on WS from the dimples as $\textbf{v}\textbf{G}_1$ approaches $\omega_{G_1}$. Here, $\omega_q= \sqrt{q^3\alpha/\rho}$ is the dispersion law for ripplons, where $\alpha$ and $\rho$ are the surface tension and the liquid density, respectively. Thus, the theory could not account for the effect of sliding observed at high driving fields. A simplified classical model was proposed by Vinen to account for both the deepening of dimples by BC scattering of ripplons, as well as sliding of WS off the dimples.~\cite{Vine1999} The model considered an infinitely long periodic (in $x$-direction) electron system moving parallel to the surface with velocity $\upsilon_x$ and subject to boundary conditions at the surface of an incompressible liquid. To account for energy losses in the system, Vinen introduced a phenomenological damping coefficient to account for natural damping of ripplons emitted by the electron lattice. He also noted that in a system of finite size, an additional contribution to damping should come from the radiative losses of ripplons through the system boundary. Damping causes a phase lag between the moving electron lattice and commensurate periodic dimple lattice, and the electron lattice slides when the driving force exceeds the maximum reaction force exerted on electrons by the dimples. The maximum force obtained by Vinen is given by

\begin{equation}
F_\textrm{max} = \frac{n_se^2E_{\perp}^2}{\rho \upsilon_d \upsilon_1},
\label{eq:FmaxVin}
\end{equation} 

\noindent where $n_s$ is the surface density of electrons, $e>0$ is the electron charge, $E_{\perp}$ is the total pressing electric field exerted on electrons perpendicular to the surface, $\upsilon_1=\sqrt{\alpha G_1/\rho}$, and $\upsilon_d$ is the phenomenological damping constant introduced by Vinen. The model predicts that before decoupling from the dimples occurs, the lattice velocity $\upsilon_x$ approaches $\upsilon_1$. Also, Eq.~(\ref{eq:FmaxVin}) predicts strong dependence of the sliding threshold on the damping of ripplons. More rigorous theoretical studies were carried out by Monarkha and Kono, who took a proper form of the electron-ripplon interaction Hamiltonian, as well as included effects of ac driving electric field typically used in the experiments.~\cite{Mona2009,Konobook} Like Vinen, they considered an infinitely extended spatially uniform WS and accounted for the natural damping of ripplons in superfluid $^4$He due to their interaction with phonons in the bulk liquid.~\cite{Roch1996} A more realistic case of an electron crystal of finite size was not considered.       

Studies of decoupling between driven WS and a liquid substrate, in particular the maximum force sustained by WS from the dimples, see Eq.~(\ref{eq:FmaxVin}), can provide important information about the general mechanism of friction.~\cite{Vano2007,Wang2008,Vano2013} Of particular interest is the study of WS of finite size, usually realized in the experiments. However, experimental study of the sliding threshold of WS and direct comparison with theoretical predictions present a challenging problem. Original experiments on the BC scattering of ripplons and sliding of WS were performed in a circular Corbino geometry in the presence of a magnetic field applied perpendicular to the surface, which significantly complicates their analysis. Recently, it was demonstrated that confining electrons in capillary-condensed microchannel devices provides many advantages for experimental studies of such charged systems.~\cite{Glas2001,Brad2011,Rees2011,Ikeg2012,Badr2016} In particular, the sliding of WS has been observed and studied in such devices without a magnetic field under either ac or dc driving electrical force.~\cite{Ikeg2009,Rees2016}

Here, we describe a new experimental study of the BC scattering of ripplons and sliding of WS on the surface of superfluid $^4$He confined in a $10$~$\mu$m-wide channel in which the effective length of the WS could be varied from 10 to 215 $\mu$m. We observed independence of the threshold driving electric field at the onset of sliding when the length of the WS exceeds 25~$\mu$m, while the threshold field strongly reduces for smaller WS. We interpret this dependence as an interplay between contributions to the loss of coherently-emitted ripplons due to, on the one hand, their damping by interaction with bulk excitations in helium and, on the other hand, their radiative loss from the finite-size WS through the system boundary. In order to quantitatively account for the experimental observations, we extended the Vinen's treatment to the case of an electron crystal of finite size. Our model allows us, in particular, to estimate the rate of natural damping of ripplons, which turned out to be consistent with available experimental data obtained by other methods.

\section{Experiment}

\begin{figure}[htt]
\includegraphics[width=0.48\textwidth]{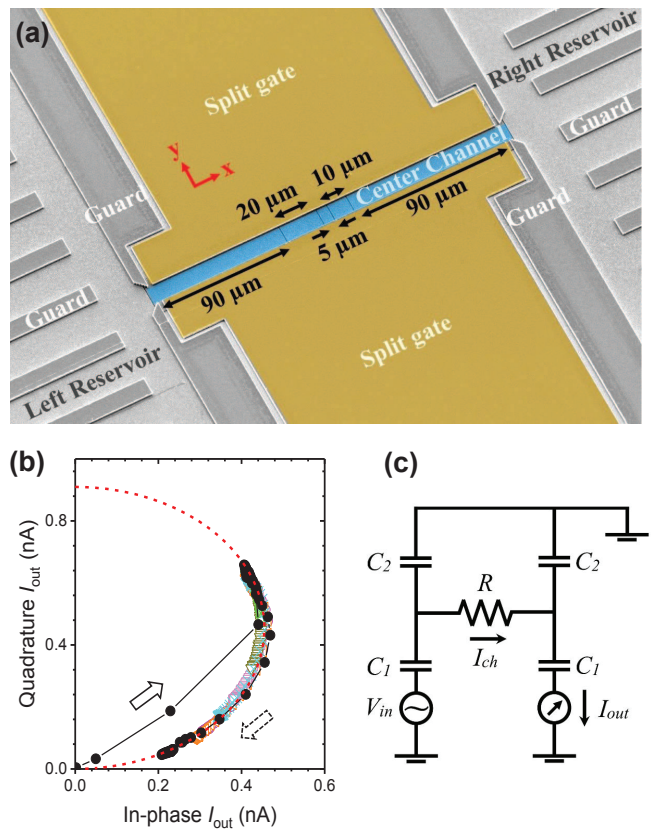}
\caption{(color online) (a) False-color scanning electron microscopic image of Sample 1 consisting of two reservoirs connected by a 215-$\mu$m channel and filled with superfluid $^4$He. The bottom electrode of the channel is segmented into pieces of different length as indicated in the Figure. (b) In-phase and quadrature components of the measured current $I_\text{out}$ due to electrons in the channel for different values of channel voltage $V_\textrm{ch}$ (solid circles connected by lines). Arrows indicate variation of $I_\text{out}$ as $V_\textrm{ch}$ increases from zero to 2~V. Other symbols plot measured components of $I_\textrm{out}$ when the voltage applied to a group of segments $V_\textrm{tr}$ is varied from 0.3 to 2~V, while the rest of segments is kept at $0.3$~V. Dashed line (red) is the fitting to the date using the lumped-circuit model shown in Fig.~\ref{fig:0}(c). (c) Electrical lumped-circuit model of the device used for the analysis. $R$ represents the resistance of the electron system which mostly comes from the resistance of the electrons in the channel.}
\label{fig:0}
\end{figure}

Two microchannel devices were used in the experiments described here. Each device was fabricated on a silicon-oxide substrate using optical lithography. The first device, hereafter called  Sample 1, was composed of two patterned gold layers separated by an insulating layer of hard-baked photoresist having thickness of about 1.6~$\mu$m.  The second device, Sample 2, was composed of similar patterned gold layers, but separated by an insulating silicon nitride layer with thickness of about 1.5~$\mu$m. Each device consisted of two sets of 20~$\mu$m-wide channel arrays connected in parallel, which served as electron reservoirs, and a 215~$\mu$m-long and 10~$\mu$m-wide central channel connecting the two reservoirs, see Fig.~\ref{fig:0}(a). The bottom gold layer consisted of three electrodes which defined the bottoms of two reservoirs and the central channel. In turn, the central channel electrode was formed by one 5~$\mu$m, one 10~ $\mu$m, one 20~$\mu$m, and two 90~$\mu$m-long segments as shown in Fg.~\ref{fig:0}(a). Adjacent electrodes were separated by 1~$\mu$m gaps. The top gold layer consisted of just two electrodes, the split-gate, and guard electrodes. The height of the channels in both devices was determined by thickness of the corresponding insulating layer. The channels were filled with superfluid $^4$He by capillary action from bulk liquid helium, the level of which was maintained slightly below the sample. 

The electrical potential at the uncharged surface of liquid helium in different parts of the microchannel device could be defined by applying independent electrical biases to different electrodes of the bottom and top layers. The surface of liquid helium in microchannels was charged with electrons produced by thermal emission from a tungsten filament placed at a distance of a few millimeters above the device, while a negative bias $V_\textrm{g}=-0.5$~V was applied to the guard and split-gate electrodes of the top layer.  All other electrodes were kept grounded. Under these conditions, the maximum surface density of electrons in the reservoirs can be estimated using a parallel-plate capacitor approximation as $n_s=\varepsilon_0\varepsilon |V_g|/(eh)$, where $h$ is the height of liquid helium in the channel, $\varepsilon_0=8.85\times 10^{-12}$~F/m is the permittivity of free space, and $\varepsilon=1.056$  is the dielectric constant of liquid helium. However, in practice we find the density of electrons in the reservoir to be appreciably less, which means that the electrical potential at the charged liquid surface, $V_\textrm{e}$, is more positive than $V_\textrm{g}$. The actual value of $V_\textrm{e}$ can be determined from experimental data as described below.    

The transport of electrons through the microchannel device was measured by the standard capacitive (Sommer-Tanner) method.~\cite{Somm1971,Ikeg2009} An ac voltage $V_\textrm{in}$ at the frequency $f$ in the range 30-100~kHz was applied to one of the reservoir electrodes, while both in-phase and quadrature components of the current $I_\textrm{out}$ induced by electron motion in the other reservoir's electrode was measured with a lock-in amplifier. An exemplary set of data taken in Sample 1 at $T=0.88$~K using driving peak-to-peak voltage $V_\textrm{in}=5$~mV at 99~kHz is shown in Fig.~\ref{fig:0}(b) where by solid (black) circles we plot both components of $I_\textrm{out}$ measured for different values of positive bias $0\leq V_\textrm{ch} \leq 2$~V applied to the central channel electrode (that is applying the same potential simultaneously to all segments constituting the channel electrode). At $V_\textrm{ch}=0$, the electrical potential at the surface of liquid helium in the central channel is slightly more negative than the potential $V_\textrm{e}$ of the charged liquid surface in the reservoirs. Thus, there are no electrons in the channel and the measured current is zero. As $V_\textrm{ch}$ increases, the electrons start filling the central channel, correspondingly the current abruptly increases as indicated by the solid-line arrow in Fig.~\ref{fig:0}(b). The corresponding threshold value of $V_\textrm{ch}$ can be used to estimate the potential $V_\textrm{e}$; therefore, the density of electrons in the reservoirs using electrostatic calculations of the electrical potential profiles in the central channel and reservoirs.~\cite{Rees2012jltp,Rees2013} Similarly, one can determine the surface density of electrons in the central channel for given values of $V_\textrm{ch}$. Note that the density of electrons in the central channel increases with $V_\textrm{ch}$ and can be substantially larger than that in the reservoirs. Correspondingly, as $V_\textrm{ch}$ increases, the amplitude of $I_\textrm{out}$ first increases and then decreases, as indicated by the dashed-line arrow in Fig.~\ref{fig:0}(b). The observed decrease of $I_\textrm{out}$ at large values of $V_\textrm{ch} \gtrsim 1$~V is due to formation of WS in the central channel for sufficiently large density of electrons, therefore causing a significant increase of electrical resistance $R$ of electrons in the channel. In order to retrieve numerical vales of $R$ from the data, we use a standard lumped-circuit model, shown in Fig.~\ref{fig:0}(c).~\cite{Ikeg2009,Ikeg2015} Here $C_1$ ($C_2$) represents capacitance between the charged liquid surface and reservoir (guard) electrodes. We assume that these capacitances are the same for both reservoirs due to symmetry of the fabricated device, while the ratio between them, $C_2/C_1$, is numerically determined using the finite element model (FEM). From this circuit, we obtain the relationship between $R$ and the amplitude of measured current $I_\textrm{out}$

\begin{equation}
R=\frac{1}{2\pi f C_0}\sqrt{\frac{V_{in}^{2}(2\pi f) ^{2} C_0^{2}\beta^{2} }{I_\text{out}^{2}}-4},
\label{eq:LumpR}
\end{equation} 

\noindent where $C_0=C_1+C_2$ and $\beta = C_1/C_0$. The value of $C_0$, which is determined by the geometry of the sample, is used as an adjustable parameter in the numerical fitting of data. An example of such a fitting with $C_0=2.87$~pF is shown in Fig.~\ref{fig:0}(b). Once this fitting parameter is determined, the value of $R$ can be calculated from the measured current $I_\textrm{out}$ using Eq.~(\ref{eq:LumpR}).

The experiment described above could be repeated by varying a positive bias $V_\textrm{tr}$ applied only to a certain group of adjacent segments that comprise the central channel electrode, while keeping the rest of segments at a small positive potential (typically 0.3~V). In this case, at sufficiently large $V_\textrm{tr}$ we observed formation of WS of the effective length corresponding to the total length of the adjacent segments biased by the potential $V_\textrm{tr}$, while electrons above the rest of the segments in the central channel were in the liquid phase. Using a segmented channel electrode, shown in Fig.~\ref{fig:0}(a), we thus could form WS of effective lengths of 5, 10, 15, 25, 35, 90, 100, 110, 120, 125, and 215~$\mu$m. The current $I_\textrm{out}$ measured by varying $V_\textrm{tr}$ from 0.3 to 2~V for different groups of segments is plotted in Fig.~\ref{fig:0}(b) using different symbols. All data fall on the same fitting curve (dashed line, red), which confirms the accuracy of our lumped-circuit model. After confirming formation of WS of adjustable length in the central channel, we proceeded to the main experiment in which we measured the $IV$-dependence for WS of different lengths by varying the amplitude of driving voltage $V_\textrm{in}$. The results and discussion are presented in the following sections. Here we note that the $IV$-curves obtained for the smallest length  (5~$\mu$m) accessible in our devices, did not exhibit well-pronounced features on nonlinear transport, such as BC plateau and sliding; therefore they are omitted from the discussion below. Indeed, FEM calculations show that WS inhomogeneity of the electron density profile can be comparable with the crystal size, which renders the above features unobservable. 

From the lumped-circuit model, the current of electrons in the central channel $I_\textrm{ch}$ is related to the measured current $I_\textrm{out}$ by $I_\textrm{ch}=(C_0/C_1)I_\textrm{out}=I_\textrm{out}/\beta$, see Fig.~\ref{fig:0} (c). In the BC scattering regime of WS transport, data analysis can be complicated by the fact that the output current response of electrons is nonlinear with respect to the sinusoidal input voltage drive. In this case, we take it into account by assuming that the lock-in amplifier measures the first harmonic of the distorted current response of electrons with respect to the sinusoidal drive, which allows us to determine the actual current $I_\textrm{BC}$ of electrons in the BC scattering regime. Values of $I_\textrm{BC}$ can be used for alternative estimate of the electron density of WS in the central channel in addition to the method described earlier. Indeed, assuming that in the BC scattering regime, the velocity of electrons is close to $\upsilon_1$, we can write $n_s=I_\textrm{BC}/(e\upsilon_1 w)$, where $w$ is the width of the electron system in the channel. In turn, the velocity $\upsilon_1$ of resonant ripplons is related to electron density through the magnitude of the first reciprocal vector of the hexagonal lattice of WS, $G_1=(8\pi^2 n_s/\sqrt{3})^{1/2}$, while the width of the electron system $w$ can be determined from FEM calculations. We found that estimations of electron density of WS from $I_\textrm{BC}$ differ by not more than $20\%$ from density estimations obtained from threshold values of $V_\textrm{ch}$, as described earlier.           

\section{Results}

\begin{figure}[htt]
\includegraphics[width=0.48\textwidth]{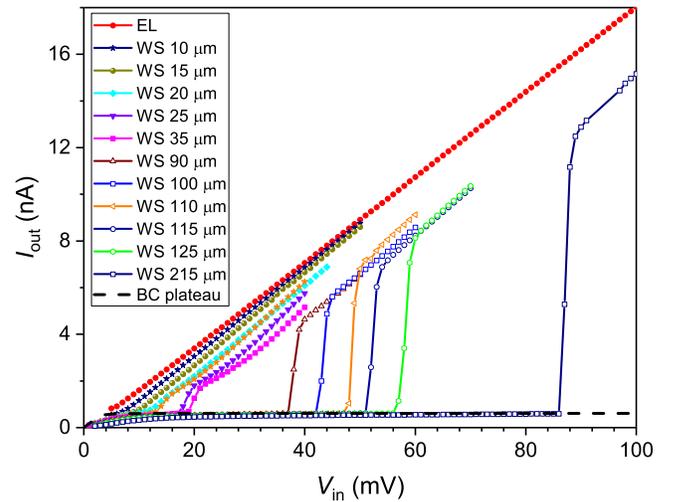}
\caption{(color online) $IV$-curves measured at $T=0.88$~K in Sample 1 for different lengths of WS in the microchannel (the corresponding length of WS is given in the legend in $\mu$m). $IV$-curve for the channel entirely filled with electron liquid is given by solid circles (EL, red). The dashed (black) line is the calculated current for WS obtained from the fitting of experimental data using the value of $I_\textrm{BC}$ as an adjustable parameter.}
\label{fig:1}
\end{figure}

Figure \ref{fig:1} shows an exemplary set of $IV$-curves taken at $T=0.88$~K using Sample 1 for different sizes of the WS formed in the microchannel by applying a positive bias $V_{\textrm{tr}}=2$~V to the corresponding segmented electrodes, while applying a positive bias of 0.3~V to the rest of the channel. As described in the previous section, in this case, an island of WS is formed only above the strongly biased electrode, while the rest of the channel is filled with the electron liquid. Both the BC plateau, as well as onset of sliding, are clearly observed for the WS when its length is at least 15~$\mu$m. The threshold driving voltage at the onset of sliding steadily increases with the size of the WS and is maximal when WS occupies the whole channel (curve marked as WS 215~$\mu$m). The dashed line (black) shows the calculated $IV$-dependence for WS obtained by fitting experimental data as described in the previous section and using the value of the current in the BC scattering regime $I_\textrm{BC}$ as the only adjustable parameter. From this fitting, we obtain $I_\textrm{BC}=0.48$~nA for the data shown in Fig.~\ref{fig:1}. This value was used to obtained an estimate for the electron density of $n_s=4.37\times 10^{13}$~cm$^{-2}$ as described in the previous section. In addition, Fig.~\ref{fig:1} shows the $IV$-curve when the entire channel is filled with the electron liquid (closed circles, red). These data were obtained by applying a uniform positive bias of $0.3$~V across the whole channel. As expected, the electron liquid shows a linear $IV$-dependence in the entire range of driving voltages. Other data sets taken with Sample 1 showed similar $IV$-curves and values of the electron density.  

\begin{figure}[htt]
\includegraphics[width=0.48\textwidth]{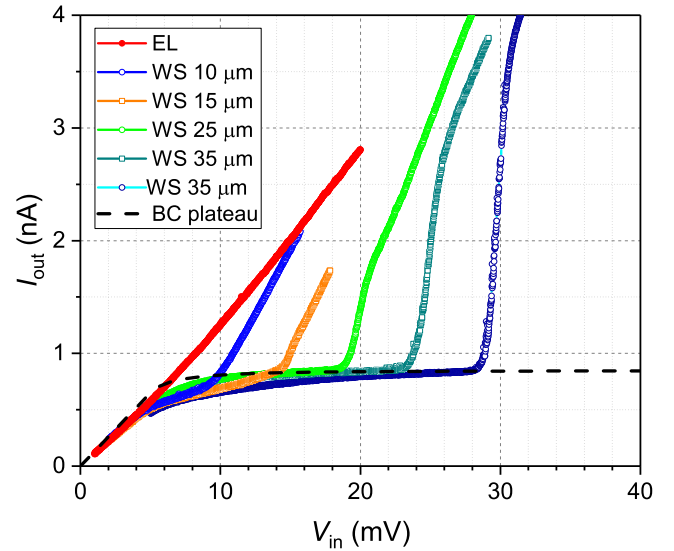}
\caption{(color online) $IV$-curves measured at $T=0.88$~K in Sample 2 for different lengths of WS in the microchannel. $IV$-curve for the channel entirely filled with electron liquid is given by solid circles (EL, red). The dashed (black) line is the calculated current for WS obtained from the fitting of experimental data using the value of $I_\textrm{BC}$ as an adjustable parameter.}
\label{fig:2}
\end{figure}

Figure \ref{fig:2} shows a set of $IV$-curves taken at $T=0.88$~K using Sample 2 for different lengths of the WS in the range from 10 to 35~$\mu$m. For this set of data, a large positive bias of $V_{\textrm{tr}}=1.5$~V was applied to the corresponding segmented electrodes to form the WS at the center of the microchannel, while keeping the rest of the channel in liquid phase. As in Fig.~\ref{fig:1}, the dashed line (black) shows the calculated current for WS obtained from the fitting of the experimental data using $I_\textrm{BC}$ as an adjustable parameter, while closed circles (red) show the $IV$-curve measured in the experiment when the entire channel is filled with the electron liquid. From the fitting we obtain $I_\textrm{BC}=0.67$~nA and the corresponding electron density $n_s=5.24\times 10^{13}$~cm$^{-2}$.          

\begin{figure}[htt]
\includegraphics[width=0.48\textwidth]{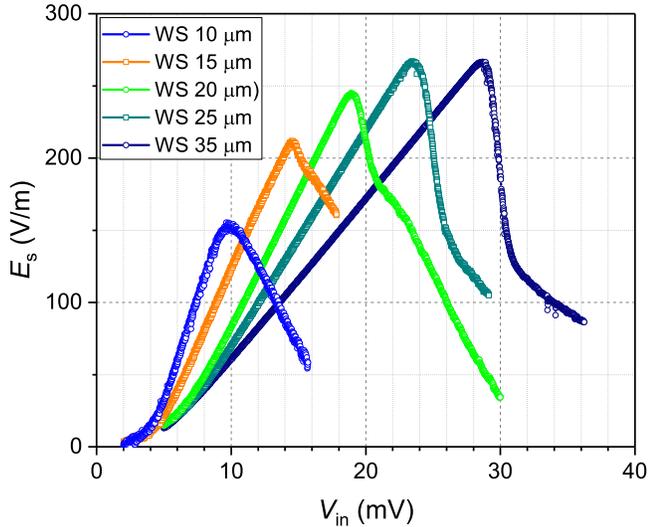}
\caption{(color online) Electric field $E_\textrm{s}$ across solid phase of the microchannel for different lengths of WS plotted as a function of driving voltage. The values of $E_\textrm{s}$ are calculated from the data in Fig.~\ref{fig:2} using Eg.~(\ref{eq:Eth}) as described in the text.}
\label{fig:3}
\end{figure}

In order to find the values for the threshold electric field $E_{\textrm{th}}$ at the onset of sliding for each length of WS, we used a simple model to account for the resistance of the micochannel filled with electrons in both solid and liquid phases. Such a model proved successful in explaining the main experimental features of electron transport in a microchannel observed in the previous experiments.~\cite{Ikeg2009} In particular, we assume that the total resistance of the microchannel, $R_\textrm{ch}$, comes from the resistance of electrons in solid and liquid phases, $R_\textrm{s}$ and $R_\textrm{l}$, respectively, which are connected in series, that is $R_\textrm{ch}=R_\textrm{s}+R_\textrm{l}$. The total resistance of the microchannel is found for a given value of the driving voltage $V_{\textrm{in}}$ using the lumped-circuit model, as described in the previous section. Then, the electric field across the WS can be estimated from the corresponding voltage drop $R_\textrm{s}I_\textrm{ch}$ according to

\begin{equation}
E_{\textrm{s}}=\frac{I_\textrm{ch}\left( R_\textrm{ch} - R_\textrm{l} \right)}{L_\textrm{tr}}, 
\label{eq:Eth}
\end{equation}     

\noindent where $I_\textrm{ch}$ is the current of electrons in the microchannel and $L_\textrm{tr}$ is the length of the strongly-biased segmented electrodes, which determine the length of the WS. The resistance $R_\textrm{l}$ is estimated as $R_\textrm{l}=R_\textrm{EL} (L_\textrm{ch}-L_\textrm{tr})/L_\textrm{ch}$, where $L_\textrm{ch}=215$~$\mu$m is the length of the microchannel and $R_\textrm{EL}$ is the resistance of the microchannel when it is entirely filled with electron liquid. Similar to $R_\textrm{ch}$, values of $R_\textrm{EL}$ were found from the lump-circuit analysis.

Figure \ref{fig:3} shows values of $E_\textrm{s}$ calculated from the data shown in Fig.~\ref{fig:2} using Eq.~(\ref{eq:Eth}). In the BC scattering regime, the electric field across the electron crystal increases linearly with the driving voltage for all WS lengths until an abrupt reduction in $E_\textrm{s}$ occurs at the onset of sliding. The latter is clearly observed in Fig.~\ref{fig:3} for all WS lengths. This determines the threshold electric field $E_\textrm{th}$. Note that after sliding, the behavior of the measured current $I_\textrm{out}$ and the electric field $E_\textrm{s}$ becomes rather complicated. At present, little is known about transport in the electron system in sliding regime. In particular, it is still under debate whether the electron system remains in the solid phase. We will not discuss this regime here.        

\begin{figure}[htt]
\includegraphics[width=0.48\textwidth]{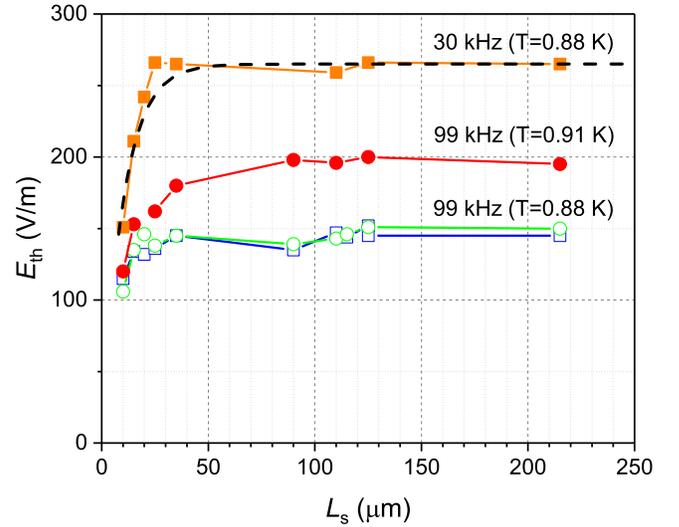}
\caption{(color online) Threshold electric field $E_\textrm{th}$ at the onset of sliding of WS plotted as a function of WS length. The values of $E_\textrm{th}$ are extracted from data taken using Sample 1 (opened symbols) and Sample 2 (closed symbols) for electron densities of $n_s=4.37\times 10^{13}$ (opened squares, blue), $4.49\times 10^{13}$ (opened circles, green), $5.24\times 10^{13}$~m$^{-2}$ (closed squares, orange) and $5.27\times 10^{13}$ (closed circles, red). The data plotted by closed (orange) squares were taken using the driving frequency $f=30$~kHz, while all other data were taken at $f=99$~kHz. In addition, the data plotted by closed (red) circles were taken at $T=0.91$~K, while all other data were taken at $T=0.88$~K. The dashed line (black) shows the dependence $E_\textrm{th}=E_0 (1-\exp(-L/L_c))$ with $E_0=265$~V/m and $L_c=10$~$\mu$m, see explanation in the text.}
\label{fig:4}
\end{figure}

Values of the threshold electric field $E_\textrm{th}$ at the onset of sliding are plotted in Fig.~\ref{fig:4} for different lengths of the WS formed in the microchannel. Data in Fig.~\ref{fig:4} show that the threshold electric field is essentially independent of the size of the WS, unless its length is shorter than about 25~$\mu$m, while for WS of shorter length there is a significant decrease of $E_\textrm{th}$. In the next section, we account for this reduction using a simple model in the spirit of Vinen's model, described in the Introduction, which incorporates the effect of finite size of the electron crystal.

\section{Discussion}

Following Vinen, we consider an essentially one-dimensional model of an electron lattice of length $L$ and periodicity $a$ moving along the microchannel at velocity $\upsilon_x$. The force exerted by electrons on the liquid surface per unit length is given by 

\begin{equation}
f(x,t)=eE_\perp\sum\limits_{n=0}^{N} \delta(x-X_n-\upsilon_xt),
\label{eq:force}
\end{equation}

\noindent where $N=L/a$ and $X_n=a n$ is the average $x$-coordinate of electrons at $t=0$. We assume that the force was averaged over the fast thermal motion of electrons; thus, the pressing field $E_\perp$ was appropriately corrected by the Debye-Waller factor~\cite{Konobook}. To proceed further, it is convenient to write the Fourier expansion of $f(x,t)$ over one-dimensional wave vectors $q$. This can easily be done by replacing the above expression for $f(x,t)$ with a similar expression as a product of an infinite train of delta-functions $\delta(x-X_n-\upsilon_xt)$, $-\infty < n <+\infty$, and a pulse function $\pi(x)=\Xi(L/2+x)-\Xi/(L/2-x)$, where $\Xi(x)$ is the Heaviside step function. This results in

\begin{equation}
f(x,t)=\frac{NeE_\perp}{\pi} \sum\limits_{m=-\infty}^{\infty} \int\limits_{-\infty}^{\infty} e^{i(qx-m\Omega t)} \frac{\sin\left[L(q-mG_1)/2\right]}{L(q-mG_1)} dq,
\label{eq:Four}
\end{equation}

\noindent where $G_1=2\pi/a$ is the first reciprocal lattice vector and $\Omega=\upsilon_x G_1$. Similar to Vinen, we consider distortion of the liquid helium surface only due to the terms $m=\pm 1$ in the above expansion, which is expected to give resonant excitation of ripplons with the one-dimensional wave vector $G_1$ when the electron velocity $\upsilon_x$ approaches the ripplon phase velocity $\upsilon_1=\sqrt{\alpha G_1/\rho}$. Higher harmonics in the expansion give resonances at higher velocities; therefore, they can be neglected. Using a boundary condition at the surface of liquid helium

\begin{equation}
-\frac{\partial p}{\partial t} + \rho \frac{\partial^2 \phi}{\partial t^2} -\alpha \frac{\partial}{\partial z} \left( \frac{\partial^2 \phi}{\partial x^2} \right)=0,
\label{eq:bound}
\end{equation} 

\noindent where $\phi$ is the velocity potential, $p=f(x,t)/w$  is the pressure on the surface from the electron lattice, and $z$-direction is perpendicular to the surface. Looking for the complex solution of the above equation in a form $\phi(x,z,t)=\int \phi_0(q) e^{qz+i(qx-\Omega t)}dq$  and using the relation between $\phi$ and the amplitude of the surface distortion in the $z$-direction $\varsigma$, $\partial \phi /\partial z = \partial \varsigma /\partial t$, we obtain

\begin{equation}
\varsigma(x,t) = \frac{2NeE_\perp}{\pi\rho w}\int\limits_{-\infty}^{\infty} \frac{qe^{i(qx-\Omega t)}}{\Omega^2 - \omega_q^2 + i\gamma_q \Omega} \frac{\sin\left[L(q-G_1)/2\right]}{L(q-G_1)} dq.
\label{eq:xi}
\end{equation}

The real part of the above equation represents the amplitude of the liquid surface deformation caused by the propagating electron lattice of length $L$. Following Vinen, we introduced a phenomenological damping rate $\gamma_q$ that accounts for natural damping of ripplons with the wave vector $q$ due to internal losses of energy in the liquid. Note that due to damping, the propagating periodic surface deformation described by the above equation has a phase lag with respect to the propagating electron lattice. That is, the positions of the minima of surface distortion do not coincide with the positions of electron lattice sites. As a result, the reaction force exerted on electrons normal to the liquid surface has a horizontal component which results in the friction force $F$ exerted on the electron system in the direction opposite to their motion. This force can be found by equating the normal component of the reaction force to $eE_\perp$, from which we obtain $F=e E_\perp \left( \partial\varsigma /\partial x \right)_{x=x_t}$, where the slope of the liquid surface in the above equation is evaluated at $x_t=\Omega t/G_1$. Plugging the real part of Eq.~(\ref{eq:xi}) into the above expression and considering the relevant wave numbers $q$ close to $G_1$, the maximum force $F$ obtained at $\upsilon_x=\upsilon_1$ can be found in the analytical form

\begin{equation}
F_\textrm{max}=\frac{n_s e^2 E_\perp^2}{\rho \upsilon_d \upsilon_1} \left[ 1 - \exp \left( -\frac{\gamma_{G_1} L}{2\upsilon_1}  \right) \right],
\label{eq:Fmax}
\end{equation}

\noindent where we introduced notation for the damping coefficient $\upsilon_d=\gamma_{G1}/G_1\approx \gamma_q/q$ which was used by Vinen. The above equation gives the maximum friction force on the electron lattice that can be provided by surface dimples. In the BC scattering regime ($\upsilon_x\approx \upsilon_1$), this force equilibrates the driving force on electrons due to applied electric field in the $x$-direction. Thus, the maximum force given by the above equation determines the threshold electric field $E_\textrm{th}$ discussed in the previous section. Note that at $L\rightarrow \infty$ the above equation recovers the Vinen's result given by Eq.~(\ref{eq:FmaxVin}) in the Introduction. More interestingly, as the length of electron lattice $L$ decreases and approaches $\upsilon_1/\gamma_{G_1}$, the maximum force and the threshold electric field $E_\textrm{th}$ also decreases. This is in agreement with our experimental observations described in the previous section. The decrease of the maximum friction force with decreasing effective size of the electron lattice has a simple physical meaning. The quantity $\upsilon_1/\gamma_{G_1}$ represents the typical propagation length of ripplons with the wave vector $G_1$ due to internal energy losses in the liquid. As long as this length is much shorter than the length of the electron lattice $L$, the damping of ripplons coherently emitted by the electron lattice does not depend on system size. On the other hand, when the propagation length becomes longer than $L$, contribution to the loss of the coherently emitted ripplons via their escape from the area occupied by the electron lattice becomes significant. This leads to weakening of dimples, and a decrease of the threshold electric field $E_\textrm{th}$, which agrees with our experimental data, shown in Fig.~\ref{fig:4}. 

To make further comparison with the experiment, we plot the dependence of $E_\textrm{th}$ on $L$ predicted by Eq.~(\ref{eq:Fmax}) in the form $E_\textrm{th}=E_0 [1-\exp(-L/L_c)]$, where $E_0$ and $L_c$ are adjustable parameters. An example of such dependence with $E_0=265$~V/cm and $L_c=10$~$\mu$m is shown in Fig.~\ref{fig:4} by a dashed line (black). Using this value of $L_c$ we can estimate the damping rate of coherently emitted ripplons under the conditions of our experiment and can compare it with available experimental data. The typical value of electron velocity in the BC scattering regime estimated in our experiment from the measured values of $I_\textrm{BC}$ and $n_s$, $\upsilon_\textrm{BC}=I_\textrm{BC}/(n_sew)$, is on the order of $10$~m/s. This gives an estimate for the velocity $\upsilon_1$. Thus, we can estimate the damping rate of ripplons $\gamma_{G_1}=2\upsilon_1/L_\textrm{c}\approx 10^6$~s$^{-1}$. The damping of micron-wavelength ripplons on the surface of superfluid $^4$He was experimentally studied by Roche {\it et al.} using an interdigital capacitor setup.~\cite{Roch1995} The authors concluded that the main contribution to damping of such ripplons comes from the ripplon-phonon interaction and provided a theoretical expression for $\gamma_q$~\cite{Roch1996}

\begin{equation}
\gamma_q=\frac{\pi^2}{90}\frac{\hbar}{\rho} \left( \frac{k_B T}{\hbar s} \right)^4 q,
\label{eq:gamq}
\end{equation}

\noindent where $s$ is the first sound velocity in liquid $^4$He. Using this expression, we obtain $\gamma_{G_1}=3\times 10^5$~s$^{-1}$ for $T=0.88$~K and $G_1=5\times 10^7$~m$^{-1}$. This agrees very satisfactorily with our order-of-magnitude estimate $\gamma_{G_1}\approx 10^6$~s$^{-1}$, considering the extreme simplicity of our model and that the theoretical formula by Roche {\it et al.} underestimates the experimentally measured attenuation coefficient at temperatures above 0.7~K.~\cite{Roch1995,Roch1996}

The treatment of an electron crystal of finite-size presented above can be incorporated with a more rigorous theoretical study of an infinitely large WS given by Monarkha and Kono.~\cite{Mona2009,Konobook} The authors defined the force acting on each electron as $-\partial \hat{V}_\textrm{int}/\partial \textbf{r}_e$ averaged over the electron distribution within the dimple and took a correct form of the electron-ripplon interaction Hamiltonian $\hat{V}_\textrm{int}$, as well as having considered electrons subject to an ac driving force. Such a treatment, although possible, is beyond the scope of this work. Here we notice that the authors predicted some broadening of BC resonances by increasing the driving frequency, which qualitatively agrees with the lower values of $E_\textrm{th}$ for higher driving frequency, as follows from data shown in Fig.~\ref{fig:4}.     

\section{Summary}

We have studied the non-linear transport of WS coupled to a commensurate deformation on the surface of liquid helium. In particular, we employed a microchannel device that allowed us to vary the effective size of the electron crystal and study its transport in a microchannel geometry. We observed dependence of the sliding threshold of the driving electric field; therefore, the maximum friction force exerted on the electron crystal from the liquid substrate, on the crystal size. In particular, we found that the friction force significantly decreases when the crystal length is shorter than about 25~$\mu$m. We explain this effect by weakening of the surface deformation due to radiative losses of ripplons coherently emitted by the driven electron lattice of finite size. To quantitatively account for the observed effect, we employed a simple hydrodynamic model that allowed us to estimate the natural damping of ripplons due to internal energy losses in the liquid. In particular, we found good agreement of our result with the predicted damping of ripplons due to their interaction with bulk excitation in liquid helium. This indicates that our experimental method is viable for studies of not only electron transport systems on liquid substrates, but also interactions between surface and bulk excitations in superfluid helium.  

{\bf Acknowledgements} The work was supported by an internal grant from Okinawa Institute of Science and Technology (OIST) Graduate University. A. O. B. was partially supported by JSPS KAKENHI Grant Number JP18K13506.





\bibliography{SCbib}

\begin{thebibliography}{34}
\expandafter\ifx\csname natexlab\endcsname\relax\def\natexlab#1{#1}\fi
\expandafter\ifx\csname bibnamefont\endcsname\relax
  \def\bibnamefont#1{#1}\fi
\expandafter\ifx\csname bibfnamefont\endcsname\relax
  \def\bibfnamefont#1{#1}\fi
\expandafter\ifx\csname citenamefont\endcsname\relax
  \def\citenamefont#1{#1}\fi
\expandafter\ifx\csname url\endcsname\relax
  \def\url#1{\texttt{#1}}\fi
\expandafter\ifx\csname urlprefix\endcsname\relax\def\urlprefix{URL }\fi
\providecommand{\bibinfo}[2]{#2}
\providecommand{\eprint}[2][]{\url{#2}}

\bibitem[{\citenamefont{Andrei}(1997)}]{Andrei}
\bibinfo{author}{\bibfnamefont{E.~Y.} \bibnamefont{Andrei}},
  \emph{\bibinfo{title}{Electrons on Helium and Other Cryogenic Substrates}}
  (\bibinfo{publisher}{Kluwer Academic, Dordrecht}, \bibinfo{year}{1997}).

\bibitem[{\citenamefont{Grimes and Adams}(1979)}]{Grim1979}
\bibinfo{author}{\bibfnamefont{C.~C.} \bibnamefont{Grimes}} \bibnamefont{and}
  \bibinfo{author}{\bibfnamefont{G.}~\bibnamefont{Adams}},
  \bibinfo{journal}{Phys. Rev. Lett.} \textbf{\bibinfo{volume}{42}},
  \bibinfo{pages}{795} (\bibinfo{year}{1979}).

\bibitem[{\citenamefont{Andrei et~al.}(1988)\citenamefont{Andrei, Deville,
  Glattli, Williams, Paris, and Etienne}}]{Andr1988}
\bibinfo{author}{\bibfnamefont{E.~Y.} \bibnamefont{Andrei}},
  \bibinfo{author}{\bibfnamefont{G.}~\bibnamefont{Deville}},
  \bibinfo{author}{\bibfnamefont{D.~C.} \bibnamefont{Glattli}},
  \bibinfo{author}{\bibfnamefont{F.~I.~B.} \bibnamefont{Williams}},
  \bibinfo{author}{\bibfnamefont{E.}~\bibnamefont{Paris}}, \bibnamefont{and}
  \bibinfo{author}{\bibfnamefont{B.}~\bibnamefont{Etienne}},
  \bibinfo{journal}{Phys. Rev. Lett.} \textbf{\bibinfo{volume}{60}},
  \bibinfo{pages}{2765} (\bibinfo{year}{1988}).

\bibitem[{\citenamefont{Zhu et~al.}(2010)\citenamefont{Zhu, Chen, Jiang, Engel,
  Tsui, Pfeiffer, and West}}]{Zhu2010}
\bibinfo{author}{\bibfnamefont{H.}~\bibnamefont{Zhu}},
  \bibinfo{author}{\bibfnamefont{Y.~P.} \bibnamefont{Chen}},
  \bibinfo{author}{\bibfnamefont{P.}~\bibnamefont{Jiang}},
  \bibinfo{author}{\bibfnamefont{L.~W.} \bibnamefont{Engel}},
  \bibinfo{author}{\bibfnamefont{D.~C.} \bibnamefont{Tsui}},
  \bibinfo{author}{\bibfnamefont{L.~N.} \bibnamefont{Pfeiffer}},
  \bibnamefont{and} \bibinfo{author}{\bibfnamefont{K.~W.} \bibnamefont{West}},
  \bibinfo{journal}{Phys. Rev. Lett.} \textbf{\bibinfo{volume}{105}},
  \bibinfo{pages}{126803} (\bibinfo{year}{2010}).

\bibitem[{\citenamefont{Liu et~al.}(2014)\citenamefont{Liu, Kamburov, Hasdemir,
  Shayegan, Pfeiffer, West, and Baldwin}}]{Liu2014}
\bibinfo{author}{\bibfnamefont{Y.}~\bibnamefont{Liu}},
  \bibinfo{author}{\bibfnamefont{D.}~\bibnamefont{Kamburov}},
  \bibinfo{author}{\bibfnamefont{S.}~\bibnamefont{Hasdemir}},
  \bibinfo{author}{\bibfnamefont{M.}~\bibnamefont{Shayegan}},
  \bibinfo{author}{\bibfnamefont{L.~N.} \bibnamefont{Pfeiffer}},
  \bibinfo{author}{\bibfnamefont{K.~W.} \bibnamefont{West}}, \bibnamefont{and}
  \bibinfo{author}{\bibfnamefont{K.~W.} \bibnamefont{Baldwin}},
  \bibinfo{journal}{Phys. Rev. Lett.} \textbf{\bibinfo{volume}{113}},
  \bibinfo{pages}{246803} (\bibinfo{year}{2014}).

\bibitem[{\citenamefont{Zhang et~al.}(2014)\citenamefont{Zhang, Huang,
  Dietsche, von Klitzing, and Smet}}]{Zhan2014}
\bibinfo{author}{\bibfnamefont{D.}~\bibnamefont{Zhang}},
  \bibinfo{author}{\bibfnamefont{X.}~\bibnamefont{Huang}},
  \bibinfo{author}{\bibfnamefont{W.}~\bibnamefont{Dietsche}},
  \bibinfo{author}{\bibfnamefont{K.}~\bibnamefont{von Klitzing}},
  \bibnamefont{and} \bibinfo{author}{\bibfnamefont{J.~H.} \bibnamefont{Smet}},
  \bibinfo{journal}{Phys. Rev. Lett.} \textbf{\bibinfo{volume}{113}},
  \bibinfo{pages}{076804} (\bibinfo{year}{2014}).

\bibitem[{\citenamefont{Murray and Winkle}(1987)}]{Murr1987}
\bibinfo{author}{\bibfnamefont{C.~A.} \bibnamefont{Murray}} \bibnamefont{and}
  \bibinfo{author}{\bibfnamefont{D.~H.~V.} \bibnamefont{Winkle}},
  \bibinfo{journal}{Phys. Rev. Lett.} \textbf{\bibinfo{volume}{58}},
  \bibinfo{pages}{1200} (\bibinfo{year}{1987}).

\bibitem[{\citenamefont{Murray and Wenk}(1989)}]{Murr1989}
\bibinfo{author}{\bibfnamefont{C.~A.} \bibnamefont{Murray}} \bibnamefont{and}
  \bibinfo{author}{\bibfnamefont{R.~A.} \bibnamefont{Wenk}},
  \bibinfo{journal}{Phys. Rev. Lett.} \textbf{\bibinfo{volume}{62}},
  \bibinfo{pages}{1643} (\bibinfo{year}{1989}).

\bibitem[{\citenamefont{Murray et~al.}(1990)\citenamefont{Murray, Sprenger, and
  Wenk}}]{Murr1990}
\bibinfo{author}{\bibfnamefont{C.~A.} \bibnamefont{Murray}},
  \bibinfo{author}{\bibfnamefont{W.~O.} \bibnamefont{Sprenger}},
  \bibnamefont{and} \bibinfo{author}{\bibfnamefont{R.~A.} \bibnamefont{Wenk}},
  \bibinfo{journal}{Phys. Rev. B} \textbf{\bibinfo{volume}{42}},
  \bibinfo{pages}{688} (\bibinfo{year}{1990}).

\bibitem[{\citenamefont{Marcus and Rice}(1996)}]{Marc1996}
\bibinfo{author}{\bibfnamefont{A.~H.} \bibnamefont{Marcus}} \bibnamefont{and}
  \bibinfo{author}{\bibfnamefont{S.~A.} \bibnamefont{Rice}},
  \bibinfo{journal}{Phys. Rev. Lett.} \textbf{\bibinfo{volume}{77}},
  \bibinfo{pages}{2577} (\bibinfo{year}{1996}).

\bibitem[{\citenamefont{Chiang and I}(1996)}]{Chia1996}
\bibinfo{author}{\bibfnamefont{C.~H.} \bibnamefont{Chiang}} \bibnamefont{and}
  \bibinfo{author}{\bibfnamefont{L.}~\bibnamefont{I}}, \bibinfo{journal}{Phys.
  Rev. Lett.} \textbf{\bibinfo{volume}{77}}, \bibinfo{pages}{647}
  (\bibinfo{year}{1996}).

\bibitem[{\citenamefont{Vanossi et~al.}(2007)\citenamefont{Vanossi, Manini,
  Caruso, Santoro, and Tosatti}}]{Vano2007}
\bibinfo{author}{\bibfnamefont{A.}~\bibnamefont{Vanossi}},
  \bibinfo{author}{\bibfnamefont{N.}~\bibnamefont{Manini}},
  \bibinfo{author}{\bibfnamefont{F.}~\bibnamefont{Caruso}},
  \bibinfo{author}{\bibfnamefont{G.~E.} \bibnamefont{Santoro}},
  \bibnamefont{and} \bibinfo{author}{\bibfnamefont{E.}~\bibnamefont{Tosatti}},
  \bibinfo{journal}{Phys. Rev. Lett.} \textbf{\bibinfo{volume}{99}},
  \bibinfo{pages}{206101} (\bibinfo{year}{2007}).

\bibitem[{\citenamefont{Wang et~al.}(2008)\citenamefont{Wang, Duang, Hong, and
  Chen}}]{Wang2008}
\bibinfo{author}{\bibfnamefont{C.-L.} \bibnamefont{Wang}},
  \bibinfo{author}{\bibfnamefont{W.-S.} \bibnamefont{Duang}},
  \bibinfo{author}{\bibfnamefont{X.-R.} \bibnamefont{Hong}}, \bibnamefont{and}
  \bibinfo{author}{\bibfnamefont{J.-M.} \bibnamefont{Chen}},
  \bibinfo{journal}{Appl. Phys. Lett.} \textbf{\bibinfo{volume}{93}},
  \bibinfo{pages}{153116} (\bibinfo{year}{2008}).

\bibitem[{\citenamefont{Vanossi et~al.}(2013)\citenamefont{Vanossi, Manini,
  Urbakh, Zapperi, and Tosatti}}]{Vano2013}
\bibinfo{author}{\bibfnamefont{A.}~\bibnamefont{Vanossi}},
  \bibinfo{author}{\bibfnamefont{N.}~\bibnamefont{Manini}},
  \bibinfo{author}{\bibfnamefont{M.}~\bibnamefont{Urbakh}},
  \bibinfo{author}{\bibfnamefont{S.}~\bibnamefont{Zapperi}}, \bibnamefont{and}
  \bibinfo{author}{\bibfnamefont{E.}~\bibnamefont{Tosatti}},
  \bibinfo{journal}{Rev. Mod. Phys.} \textbf{\bibinfo{volume}{85}},
  \bibinfo{pages}{529} (\bibinfo{year}{2013}).

\bibitem[{\citenamefont{Bylinskii et~al.}(2015)\citenamefont{Bylinskii,
  Gangloff, and Vuletic}}]{Byli2015}
\bibinfo{author}{\bibfnamefont{A.}~\bibnamefont{Bylinskii}},
  \bibinfo{author}{\bibfnamefont{D.}~\bibnamefont{Gangloff}}, \bibnamefont{and}
  \bibinfo{author}{\bibfnamefont{V.}~\bibnamefont{Vuletic}},
  \bibinfo{journal}{Science} \textbf{\bibinfo{volume}{348}},
  \bibinfo{pages}{1115} (\bibinfo{year}{2015}).

\bibitem[{\citenamefont{Kristensen et~al.}(1996)\citenamefont{Kristensen,
  Djerfi, Fozooni, Lea, Richardson, Santrich-Badal, Blackburn, and van~der
  Heijden}}]{Kris1996}
\bibinfo{author}{\bibfnamefont{A.}~\bibnamefont{Kristensen}},
  \bibinfo{author}{\bibfnamefont{K.}~\bibnamefont{Djerfi}},
  \bibinfo{author}{\bibfnamefont{P.}~\bibnamefont{Fozooni}},
  \bibinfo{author}{\bibfnamefont{M.~J.} \bibnamefont{Lea}},
  \bibinfo{author}{\bibfnamefont{P.~J.} \bibnamefont{Richardson}},
  \bibinfo{author}{\bibfnamefont{A.}~\bibnamefont{Santrich-Badal}},
  \bibinfo{author}{\bibfnamefont{A.}~\bibnamefont{Blackburn}},
  \bibnamefont{and} \bibinfo{author}{\bibfnamefont{R.~W.} \bibnamefont{van~der
  Heijden}}, \bibinfo{journal}{Phys. Rev. Lett.} \textbf{\bibinfo{volume}{77}},
  \bibinfo{pages}{1350} (\bibinfo{year}{1996}).

\bibitem[{\citenamefont{Ikegami et~al.}(2009)\citenamefont{Ikegami, Akimoto,
  and Kono}}]{Ikeg2009}
\bibinfo{author}{\bibfnamefont{H.}~\bibnamefont{Ikegami}},
  \bibinfo{author}{\bibfnamefont{H.}~\bibnamefont{Akimoto}}, \bibnamefont{and}
  \bibinfo{author}{\bibfnamefont{K.}~\bibnamefont{Kono}},
  \bibinfo{journal}{Phys. Rev. Lett.} \textbf{\bibinfo{volume}{102}},
  \bibinfo{pages}{046807} (\bibinfo{year}{2009}).

\bibitem[{\citenamefont{Shirahama and Kono}(1995)}]{Shir1995}
\bibinfo{author}{\bibfnamefont{K.}~\bibnamefont{Shirahama}} \bibnamefont{and}
  \bibinfo{author}{\bibfnamefont{K.}~\bibnamefont{Kono}},
  \bibinfo{journal}{Phys. Rev. Lett.} \textbf{\bibinfo{volume}{74}},
  \bibinfo{pages}{781} (\bibinfo{year}{1995}).

\bibitem[{\citenamefont{Rees et~al.}(2016)\citenamefont{Rees, Beysengulov, Lin,
  and Kono}}]{Rees2016}
\bibinfo{author}{\bibfnamefont{D.~G.} \bibnamefont{Rees}},
  \bibinfo{author}{\bibfnamefont{N.~R.} \bibnamefont{Beysengulov}},
  \bibinfo{author}{\bibfnamefont{J.~J.} \bibnamefont{Lin}}, \bibnamefont{and}
  \bibinfo{author}{\bibfnamefont{K.}~\bibnamefont{Kono}},
  \bibinfo{journal}{Phys. Rev. Lett.} \textbf{\bibinfo{volume}{116}},
  \bibinfo{pages}{206801} (\bibinfo{year}{2016}).

\bibitem[{\citenamefont{Dykman and Rubo}(1997)}]{Dykm1997}
\bibinfo{author}{\bibfnamefont{M.~I.} \bibnamefont{Dykman}} \bibnamefont{and}
  \bibinfo{author}{\bibfnamefont{Y.~G.} \bibnamefont{Rubo}},
  \bibinfo{journal}{Phys. Rev. Lett.} \textbf{\bibinfo{volume}{78}},
  \bibinfo{pages}{4813} (\bibinfo{year}{1997}).

\bibitem[{\citenamefont{Vinen}(1999)}]{Vine1999}
\bibinfo{author}{\bibfnamefont{W.~F.} \bibnamefont{Vinen}},
  \bibinfo{journal}{J. Phys.: Condens. Matter} \textbf{\bibinfo{volume}{11}},
  \bibinfo{pages}{9709} (\bibinfo{year}{1999}).

\bibitem[{\citenamefont{Monarkha and Kono}(2009)}]{Mona2009}
\bibinfo{author}{\bibfnamefont{Y.~P.} \bibnamefont{Monarkha}} \bibnamefont{and}
  \bibinfo{author}{\bibfnamefont{K.}~\bibnamefont{Kono}},
  \bibinfo{journal}{Low. Temp. Phys.} \textbf{\bibinfo{volume}{35}},
  \bibinfo{pages}{356} (\bibinfo{year}{2009}).

\bibitem[{\citenamefont{Monarkha and Kono}(2004)}]{Konobook}
\bibinfo{author}{\bibfnamefont{Y.~P.} \bibnamefont{Monarkha}} \bibnamefont{and}
  \bibinfo{author}{\bibfnamefont{K.}~\bibnamefont{Kono}},
  \emph{\bibinfo{title}{Two-Dimensional Coulomb Liquids and Solids}}
  (\bibinfo{publisher}{Springer-Verlag, Berlin}, \bibinfo{year}{2004}).

\bibitem[{\citenamefont{Roche et~al.}(1996)\citenamefont{Roche, Roger, and
  Williams}}]{Roch1996}
\bibinfo{author}{\bibfnamefont{P.}~\bibnamefont{Roche}},
  \bibinfo{author}{\bibfnamefont{M.}~\bibnamefont{Roger}}, \bibnamefont{and}
  \bibinfo{author}{\bibfnamefont{F.~I.~B.} \bibnamefont{Williams}},
  \bibinfo{journal}{Phys. Rev. B} \textbf{\bibinfo{volume}{53}},
  \bibinfo{pages}{2225} (\bibinfo{year}{1996}).

\bibitem[{\citenamefont{Glasson et~al.}(2001)\citenamefont{Glasson, Dotsenko,
  Fozooni, Lea, Bailey, Papageorgiou, Andresen, and Kristensen}}]{Glas2001}
\bibinfo{author}{\bibfnamefont{P.}~\bibnamefont{Glasson}},
  \bibinfo{author}{\bibfnamefont{V.}~\bibnamefont{Dotsenko}},
  \bibinfo{author}{\bibfnamefont{P.}~\bibnamefont{Fozooni}},
  \bibinfo{author}{\bibfnamefont{M.~J.} \bibnamefont{Lea}},
  \bibinfo{author}{\bibfnamefont{W.}~\bibnamefont{Bailey}},
  \bibinfo{author}{\bibfnamefont{G.}~\bibnamefont{Papageorgiou}},
  \bibinfo{author}{\bibfnamefont{S.~E.} \bibnamefont{Andresen}},
  \bibnamefont{and}
  \bibinfo{author}{\bibfnamefont{A.}~\bibnamefont{Kristensen}},
  \bibinfo{journal}{Phys. Rev. Lett.} \textbf{\bibinfo{volume}{87}},
  \bibinfo{pages}{176802} (\bibinfo{year}{2001}).

\bibitem[{\citenamefont{Bradbury et~al.}(2011)\citenamefont{Bradbury, Takita,
  Gurrieri, Wilkel, Eng, Carroll, and Lyon}}]{Brad2011}
\bibinfo{author}{\bibfnamefont{F.~R.} \bibnamefont{Bradbury}},
  \bibinfo{author}{\bibfnamefont{M.}~\bibnamefont{Takita}},
  \bibinfo{author}{\bibfnamefont{T.~M.} \bibnamefont{Gurrieri}},
  \bibinfo{author}{\bibfnamefont{K.~J.} \bibnamefont{Wilkel}},
  \bibinfo{author}{\bibfnamefont{K.}~\bibnamefont{Eng}},
  \bibinfo{author}{\bibfnamefont{M.~S.} \bibnamefont{Carroll}},
  \bibnamefont{and} \bibinfo{author}{\bibfnamefont{S.~A.} \bibnamefont{Lyon}},
  \bibinfo{journal}{Phys. Rev. Lett.} \textbf{\bibinfo{volume}{107}},
  \bibinfo{pages}{266803} (\bibinfo{year}{2011}).

\bibitem[{\citenamefont{Rees et~al.}(2011)\citenamefont{Rees, Kuroda,
  Marrache-Kikuchi, Hofer, Leiderer, and Kono}}]{Rees2011}
\bibinfo{author}{\bibfnamefont{D.~G.} \bibnamefont{Rees}},
  \bibinfo{author}{\bibfnamefont{I.}~\bibnamefont{Kuroda}},
  \bibinfo{author}{\bibfnamefont{C.~A.} \bibnamefont{Marrache-Kikuchi}},
  \bibinfo{author}{\bibfnamefont{M.}~\bibnamefont{Hofer}},
  \bibinfo{author}{\bibfnamefont{P.}~\bibnamefont{Leiderer}}, \bibnamefont{and}
  \bibinfo{author}{\bibfnamefont{K.}~\bibnamefont{Kono}},
  \bibinfo{journal}{Phys. Rev. Lett} \textbf{\bibinfo{volume}{106}},
  \bibinfo{pages}{026803} (\bibinfo{year}{2011}).

\bibitem[{\citenamefont{Ikegami et~al.}(2012)\citenamefont{Ikegami, Akimoto,
  Rees, and Kono}}]{Ikeg2012}
\bibinfo{author}{\bibfnamefont{H.}~\bibnamefont{Ikegami}},
  \bibinfo{author}{\bibfnamefont{H.}~\bibnamefont{Akimoto}},
  \bibinfo{author}{\bibfnamefont{D.~G.} \bibnamefont{Rees}}, \bibnamefont{and}
  \bibinfo{author}{\bibfnamefont{K.}~\bibnamefont{Kono}},
  \bibinfo{journal}{Phys. Rev. Lett.} \textbf{\bibinfo{volume}{109}},
  \bibinfo{pages}{236802} (\bibinfo{year}{2012}).

\bibitem[{\citenamefont{Badrutdinov et~al.}(2016)\citenamefont{Badrutdinov,
  Smorodine, Rees, Lin, and Konstantinov}}]{Badr2016}
\bibinfo{author}{\bibfnamefont{A.~O.} \bibnamefont{Badrutdinov}},
  \bibinfo{author}{\bibfnamefont{A.~V.} \bibnamefont{Smorodine}},
  \bibinfo{author}{\bibfnamefont{D.~G.} \bibnamefont{Rees}},
  \bibinfo{author}{\bibfnamefont{J.~Y.} \bibnamefont{Lin}}, \bibnamefont{and}
  \bibinfo{author}{\bibfnamefont{D.}~\bibnamefont{Konstantinov}},
  \bibinfo{journal}{Phys. Rev. B} \textbf{\bibinfo{volume}{94}},
  \bibinfo{pages}{195311} (\bibinfo{year}{2016}).

\bibitem[{\citenamefont{Sommer and Tanner}(1971)}]{Somm1971}
\bibinfo{author}{\bibfnamefont{W.~T.} \bibnamefont{Sommer}} \bibnamefont{and}
  \bibinfo{author}{\bibfnamefont{D.~J.} \bibnamefont{Tanner}},
  \bibinfo{journal}{Phys. Rev. Lett.} \textbf{\bibinfo{volume}{27}},
  \bibinfo{pages}{1345} (\bibinfo{year}{1971}).

\bibitem[{\citenamefont{Rees et~al.}(2012)\citenamefont{Rees, Kuroda,
  Marrache-Kikuchi, Hofer, and Kono}}]{Rees2012jltp}
\bibinfo{author}{\bibfnamefont{D.~G.} \bibnamefont{Rees}},
  \bibinfo{author}{\bibfnamefont{I.}~\bibnamefont{Kuroda}},
  \bibinfo{author}{\bibfnamefont{C.~A.} \bibnamefont{Marrache-Kikuchi}},
  \bibinfo{author}{\bibfnamefont{M.}~\bibnamefont{Hofer}}, \bibnamefont{and}
  \bibinfo{author}{\bibfnamefont{K.}~\bibnamefont{Kono}}, \bibinfo{journal}{J.
  Low Temp. Phys.} \textbf{\bibinfo{volume}{166}}, \bibinfo{pages}{107}
  (\bibinfo{year}{2012}).

\bibitem[{\citenamefont{Rees et~al.}(2013)\citenamefont{Rees, Ikegami, and
  Kono}}]{Rees2013}
\bibinfo{author}{\bibfnamefont{D.~G.} \bibnamefont{Rees}},
  \bibinfo{author}{\bibfnamefont{H.}~\bibnamefont{Ikegami}}, \bibnamefont{and}
  \bibinfo{author}{\bibfnamefont{K.}~\bibnamefont{Kono}}, \bibinfo{journal}{J.
  Phys. Soc. Jpn.} \textbf{\bibinfo{volume}{82}}, \bibinfo{pages}{124602}
  (\bibinfo{year}{2013}).

\bibitem[{\citenamefont{Ikegami et~al.}(2015)\citenamefont{Ikegami, Akimoto,
  and Kono}}]{Ikeg2015}
\bibinfo{author}{\bibfnamefont{H.}~\bibnamefont{Ikegami}},
  \bibinfo{author}{\bibfnamefont{H.}~\bibnamefont{Akimoto}}, \bibnamefont{and}
  \bibinfo{author}{\bibfnamefont{K.}~\bibnamefont{Kono}}, \bibinfo{journal}{J.
  Low Temp. Phys.} \textbf{\bibinfo{volume}{179}}, \bibinfo{pages}{251}
  (\bibinfo{year}{2015}).

\bibitem[{\citenamefont{Roche et~al.}(1995)\citenamefont{Roche, Deville,
  Keshichev, Appleyard, and Williams}}]{Roch1995}
\bibinfo{author}{\bibfnamefont{P.}~\bibnamefont{Roche}},
  \bibinfo{author}{\bibfnamefont{G.}~\bibnamefont{Deville}},
  \bibinfo{author}{\bibfnamefont{K.~O.} \bibnamefont{Keshichev}},
  \bibinfo{author}{\bibfnamefont{N.~J.} \bibnamefont{Appleyard}},
  \bibnamefont{and} \bibinfo{author}{\bibfnamefont{F.~I.~B.}
  \bibnamefont{Williams}}, \bibinfo{journal}{Phys. Rev. Lett.}
  \textbf{\bibinfo{volume}{75}}, \bibinfo{pages}{3316} (\bibinfo{year}{1995}).

\end{thebibliography}

\end{document}